\documentclass[11pt]{elsart}

\usepackage{amssymb,amsmath}
\usepackage{graphicx}
\journal{Physica A}

\begin{document}
\begin{frontmatter}
\title{Patterns in hydraulic ripples \\with binary granular mixtures}
\author{H. Caps and N. Vandewalle}
\address{GRASP, Institut de Physique B5, Universit\'e de Li\`ege, 4000 Li\`ege, Belgium}

\begin{abstract}
An experimental study of a binary granular mixture submitted to a transient shear flow in a cylindrical container is reported. The formation of ripples with a spiral shape is observed. The appearance of phase segregation in those spiral patterns is shown. The relative grain size bewteen sand species is found to be a relevant parameter leading to phase segregation. However, the relative repose angle is an irrelevant parameter. \\ The formation of sedimentary structures is also presented. They result from a ripple climbing process. The ``sub-critical'' or ``super-critical''  character of the lamination patterns is shown to depend on the rotation speed of the container.
\end{abstract}
\end{frontmatter}

\section{Introduction}
The ripple formation on a sand bed eroded by a fluid, such as air or water, is one of the most famous phenomena of pattern formation \cite{ball}. Indeed, coastal areas as well as desert landscapes are covered by those structures. In spite of their familiar aspect, the involved physical mechanisms are related to complex dynamical processes of granular transport as described by Bagnold \cite{bagnold1}. 

Recently, sand ripples created by water shear flows have received much attention \cite{frette,bagnold2,stegner}. Previous experiments have been performed in rectangular \cite{bagnold2} or annular \cite{frette,stegner} channels filled with water and sand. If the fluid motion is oscillatory, the created ripples are symmetric (both ripple faces have the same length). On the contrary, the ripples created under a non-oscillating fluid motion are generally asymmetric \cite{bagnold1,goossens}. Indeed, the upstream face (stoss slope) is generally larger than the downstream face (lee slope). In coastal areas, ripples are generally asymmetric. 

Moreover, one should note that natural sand beds are composed of different granular species. Broad granulometric distributions are indeed observed. As a consequence, one can observe phase segregation and stratigraphy \cite{hunter}. 

In the present paper, we report an experimental study of a binary granular mixture submitted to a non-oscillating water shear flow. Patterns, similar to natural ones, are observed and discussed. In the next Section, the experimental setup is presented. The mechanisms of ripple formation are briefly discussed in Section 3. Phase segregation and sedimentary structures are observed and discussed in Section 4 and Section 5 respectively. Finally, a summary of our findings is given in Section 6.

\section{Experimental setup}

\begin{figure}[h]
\begin{center}
\includegraphics[width=6cm]{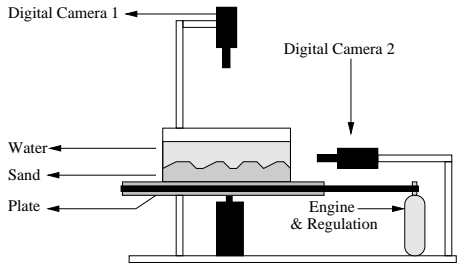}
\caption{\small Experimental setup for the ripple formation and subsequent analysis.}\label{manip}
\end{center}
\end{figure}

Our experimental setup is illustrated in Figure \ref{manip} and consists in a horizontal circular plate connected by a belt to an engine. The rotation speed can be adjusted from 8 rpm to 100 rpm. A cylindrical container (14 cm diameter) is filled with water (400 ml) and sand (220 ml). The container is placed in the center of the plate and put into rotation. When the rotation is brutally stopped, the sand bed is dragged by the water which continues its inertial circular motion. After a short time (typically 2s), ripples are observed. A CCD camera (\# 1) is placed on the top of the circular container and records top view images of the landscape. A second CCD camera (\# 2) records transversal views of the granular landscape. 

\begin{figure}[h]
\begin{center}
\includegraphics[width=8cm]{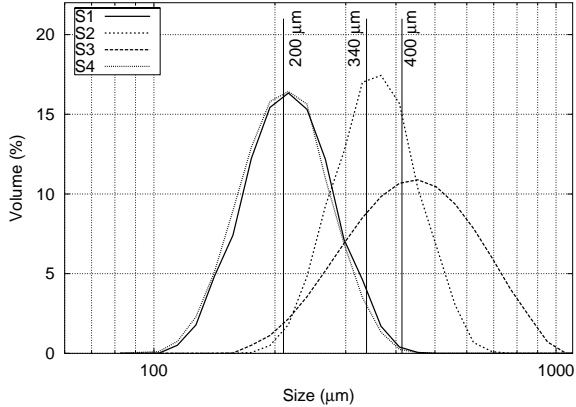}
\caption{\small Statistical distribution of the sand grain sizes. Four sand species are illustrated: white sand S1 ($d$=200 $\mu$m, $\theta_r$=33$^\circ$), green sand S2 ($d$=400 $\mu$m, $\theta_r$=33$^\circ$), black sand S3 ($d$=340 $\mu$m, $\theta_r$=28$^\circ$), and black sand S4 ($d$=200 $\mu$m, $\theta_r$=28$^\circ$).}\label{distrib}
\end{center}
\end{figure}

We have performed several experiments with equal-volume binary mixtures. Sand species ($S_i$) are different in color, mean grain diameter $d$, and repose angle $\theta_r$. We have used black, white and green sands sifted such that they have different granulometric distributions. Those granulometric distributions are presented in Figure \ref{distrib}. Grain sizes are log-normaly distributed around the means $d$=200 $\mu$m and $d$=340 $\mu$m for the black sand, $d$=400 $\mu$m for the green one, and $d$=200 $\mu$m for the white one. The white (S1) and green (S2) sands have repose angles $\theta_r=33^\circ$, while the black one (S3, S4)  has a repose angle value of $\theta_r=28^\circ$. Thus, S1 and S4 species have the same size distribution but different repose angles. Finally, one should note that all sand species have similar spherical-like shapes at the microscopic scale.

Sand mixtures have been prepared from S1,$\ldots$, S4 in order to check the influence of both $d$ and $\theta_r$ parameters on the ripple morphology. We have made 4 different types of binary mixtures: (i) grain species differing in size but having the same repose angle, (ii) grain species differing in repose angle but having the same size, (iii) mixture with the largest grain species having the smallest repose angle, and (iv) mixture with the largest grain species having the largest angle of repose. Sand species are defined by the couple of intrinsic sand parameters $(d,\theta)$. 

\section{Ripple formation}

An optically flat granular surface observed at the grain level is nevertheless rough. As a consequence, when a fluid flows over this kind of surface, a turbulent boundary layer is developed at large Reynolds number. The fluid speed at a given height $z$ can be written according to the Prandtl-von K\'arm\'an law \cite{engelund}:
\begin{equation}
u(z)=u_0+\frac{u_\star}{K}\ln\left(\frac{z}{z_0}\right),
\end{equation}   
where $u_0$ is the fluid velocity at the reference height $z_0$, chosen for the best convenience. Usually, $z_0$ is chosen to be $z_0\simeq d/30$ for a granular sand surface composed of grains with diameter $d$. The parameter $K$ is called the von K\'arm\'an constant and has to be calculated experimentally \cite{schlichting}, while $u_\star$ is the shear velocity, related to the shear stress $\tau$ by
\begin{equation}
\tau=\rho_f\,u_\star,
\end{equation} 
where $\rho_f$ is the fluid density. 

Since the fluid speed depends on the vertical coordinate $z$, sand grains are submitted to a vertical velocity gradient. Underlying drag and buoyant forces lead to the carrying of grains in the fluid. Three possible grain motions are recognized \cite{bagnold1}: saltation, reptation and suspension. Due to this granular transport, the surface becomes irregular. The largest irregularities are nucleation centers of the ripples. One should note that in our experiment, the rotation speed $\omega$ of the container controls the direction and amplitude of the velocity gradient mentioned above.

\begin{figure}[h]
\begin{center}
\includegraphics[width=10cm]{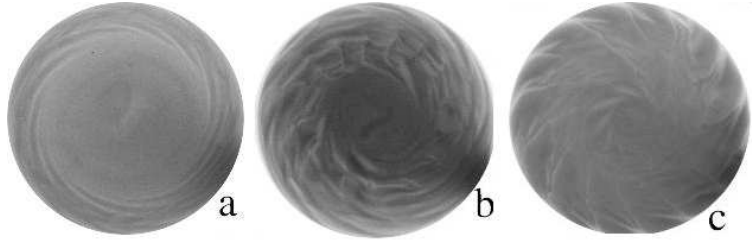}
\caption{Depending on the rotation speed, three patterns can be observed: (a) at low rotation speeds ($\omega\leq 30$ rpm) the ripples forms one spiral, (b) at medium speeds ($30<\omega<45$ rpm) two spirals are superposed, one is more open than the other one, (c) at high rotation speeds ($\omega\geq 45$ rpm) the two spirals meet each others at a given radius.}\label{phases}
\end{center}
\end{figure}

Figure \ref{phases} presents top views of the whole landscape after ripple formation in the case of a monodisperse sand bed. Three rotation speeds are illustrated: $\omega=25$ rpm, $\omega=40$ rpm and $\omega=55$ rpm. One should note the spiral shape formed by the crests of the ripples. Actually, ripples constitute the arms of the spiral. Such spirals should be related to an instability (such as Tollmien-Schlichting waves \cite{schlichting}) in the fluid motion. As granular transport directly depends on the fluid motion, the subsequent ripples keep the print of that instability. The same kind of patterns has been previously observed in the boundary layer of a fluid rotating over a stainless-steel disk \cite{jarre,gauthier}. Those studies have shown that the patterns depend on the rotation speed of the fluid: (i) at low rotation speeds ($R_e<120$) the patterns are concentric circles, (ii) at medium rotation speeds ($120\leq R_e\leq 180$), spirals are created, (iii) for high rotation speeds ($R_e>300$), a disordered state is observed because turbulence takes place in the center of the disk. Similar effects have been observed in our experiments. Indeed, at low rotation speeds (typically $30$ rpm) well defined spirals are observed. When the rotation speed increases ($30\leq\omega\leq 45$ rpm), two ``superposed'' spirals are observed: one having a larger angle than the other, i.e. one having closer arms than the other. For high rotation speeds ($\omega>45$ rpm) the branches of the two spirals are coupled two by two. The detailed study of the pattern geometry is outside the scope of this work and is let for further experiments dealing with selected parameters such as the Reynolds number. Such work is in progress and will be published later \cite{spirals}. 

\section{Phase segregation}

\begin{figure}[h]
\begin{center}
\includegraphics[width=10cm]{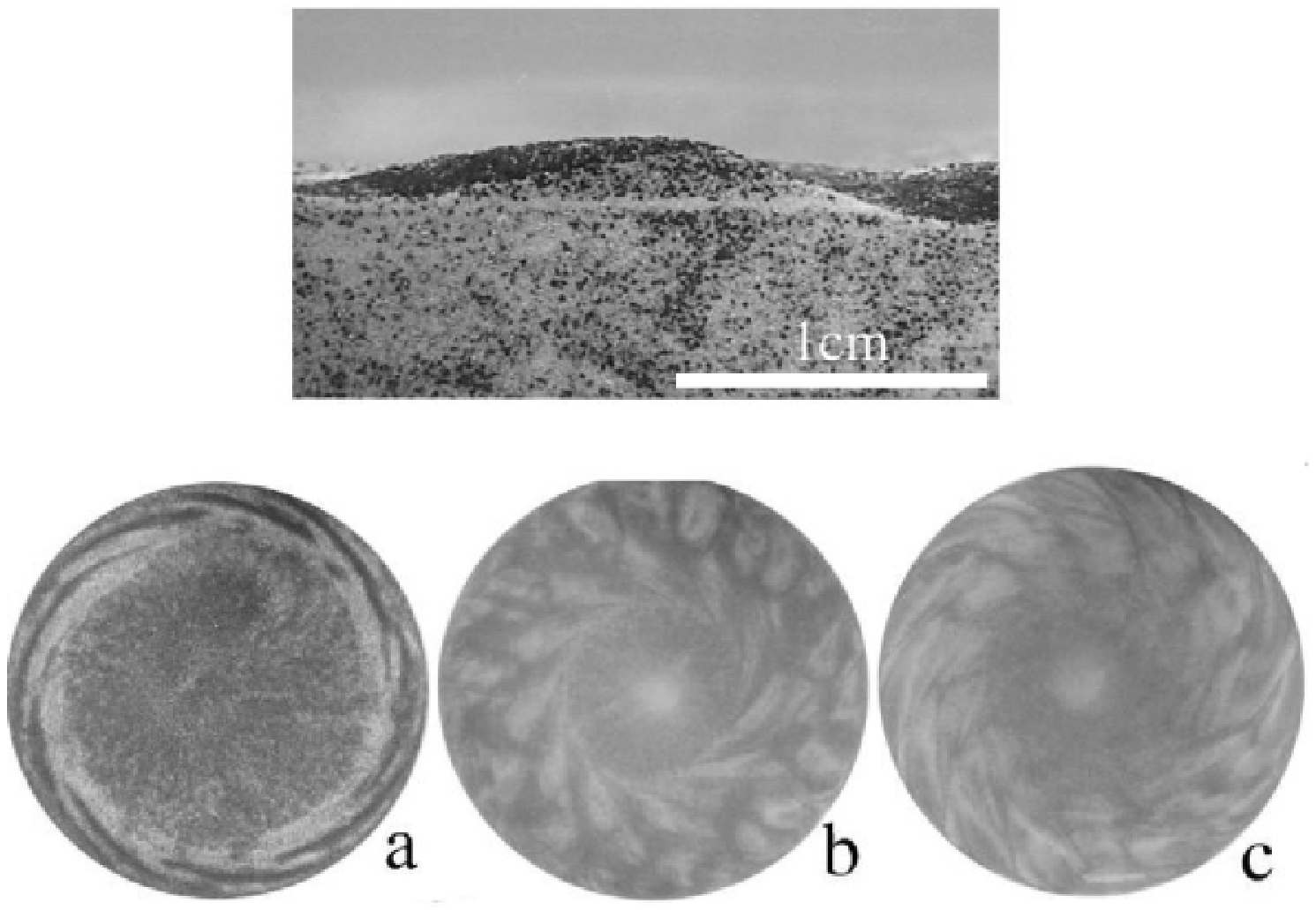}
\caption{\small (Top) Transverse view of one ripple. A phase segregation is observed: the larger grains (black) are found on the lee slope, while the smaller grains (white) mainly segregate on the stoss slope. The water flows from right to left. (Bottom) Depending on the rotation speed, three patterns can be observed: (a) one spiral, (b) two superposed spirals and (c) two spirals meeting at some radius. For all a, b and c pictures, the larger grains are found on the lee slope. The water flows clockwise and all pictures are 11 cm in diameter.}\label{bw}
\end{center}
\end{figure}

Figure \ref{bw} presents a typical result of a series of experiments with a mixture of S1 and S3 sand species. One should note a {\it phase segregation} of the sand species: the black grains ($d=340\,\mu$m) segregate on the lee slopes of the ripples, while the white grains ($d=200\,\mu$m) are found on the stoss slopes.

We have performed various experiments on all granular mixtures described above, with different rotation speeds and water volumes. We have noticed that any finite size difference between both granular species leads to a phase segregation. On the contrary, the angle of repose is not a physical parameter leading to phase segregation, even though this parameter is relevant for other segregation phenomena \cite{noe}. Moreover, it seems that the phase segregation itself does not depend on the rotation speed of the container. Indeed, larger grains have been found on the lee slopes of the ripples for all values of the rotation speed. 

Actually, the mechanism of phase segregation should be understood as follows. Depending on the fluid shear stress, i.e. on the rotation speed $\omega$, three cases can be observed: (i) at small shear stress ($\omega<30$ rpm), reptation of the smaller grains is the only granular motion observed, (ii) at medium shear stress  ($30\leq\omega\leq 45$ rpm), the small grains are carried by saltation, the large ones being transported by reptation, (iii) at high shear stress ($\omega>45$ rpm), both granular species move by saltation.

In the case of a low rotation speed, tiny ripples are created and are called rolling grain ripples \cite{frette,stegner}. Our experimental setup does not allow us to study phase segregation in such ripples.

\begin{figure}[h]
\begin{center}
\includegraphics[width=13cm]{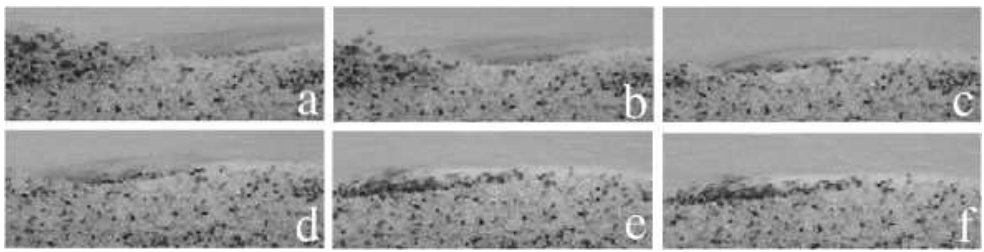}
\caption{\small Six steps in the formation of a ripple with two kinds of sand. (a) The impacts caused by the landing of the small grains (white) after their saltation jump lead to ripple formation. (b,c) The fog formed by the small grains carries the larger ones (black) in a reptation motion. (d,e) As the fluid speed vanishes, the larger grains continue their inertial reptation motion and fall in avalanches on the lee slope of the ripple. (f) The large grains are found on the lee slope, and small grains are found on the stoss slope.}\label{segrall}
\end{center}
\end{figure}

Both cases of medium and high rotation speeds will be treated together, as the observed phase segregation is similar for both cases. The process of phase segregation occurs as follows [see Figure \ref{segrall}]. Because of their weight, the smaller grains are more easily carried by the fluid than the larger ones (a). Those small grains create a saltation fog over the sand bed (b,c). Although the fluid speed is not large enough to carry the large grains in a saltation motion, the fog captures and transports them (d). As the vessel motion is stopped (e), no saltation is observed anymore after a finite time (typically 1s). Nevertheless, due to their inertia, some grains, essentially the larger ones, are obsered to move further by reptation during a short time (less than 1 s). Arriving on the crest of a ripple, they roll on the lee slope. Because of frictional forces between those grains and the static ones, they are soon stopped. As a consequence, larger grains are found near the ripple crests, while the smaller ones are mainly found near the stoss slope bottoms. 
     
\section{Sedimentation}
\begin{figure}[h]
\begin{center}
\includegraphics[width=5cm]{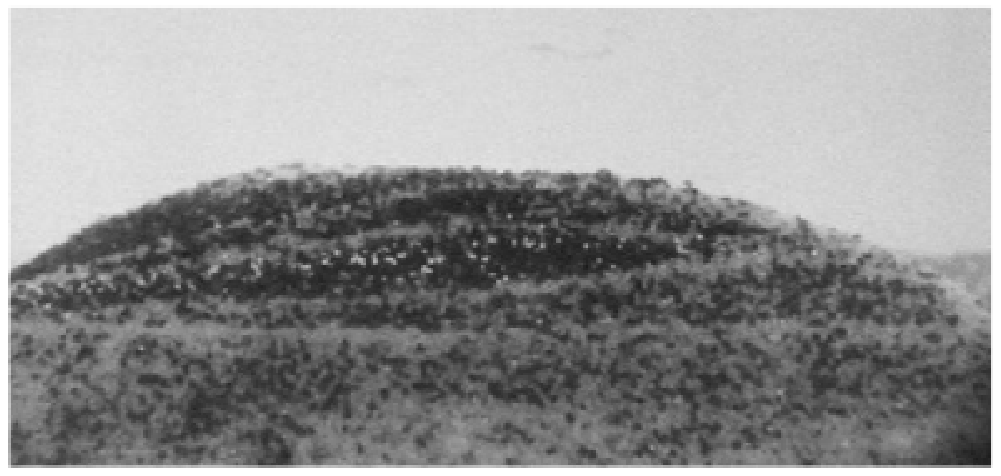}\hspace{1cm}\includegraphics[width=5cm]{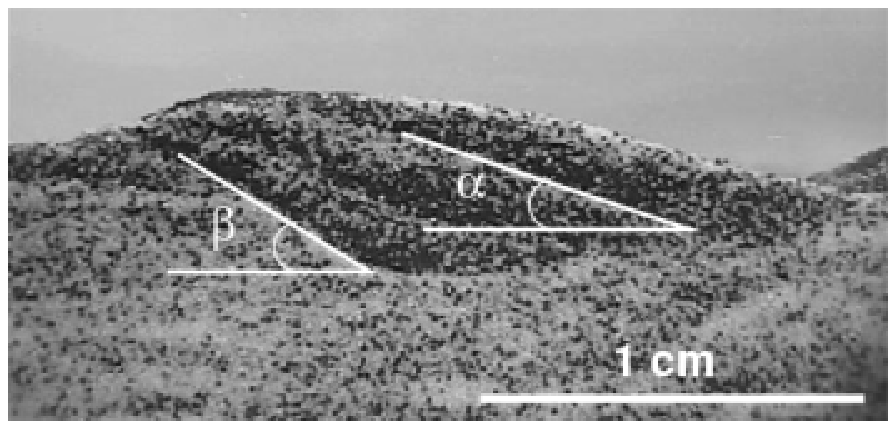}
\caption{\small Transversal view of two sedimentary structures obtained after the superposition of three ripples. (left) Medium rotation speed creates a complete rippleform lamination while (right) high rotation speed creates a truncated ripple-foreset crosslamination. Both pictures have the same scale.}\label{strat}
\end{center}
\end{figure}

Starting with a rippled sand bed, we have repeated the turn/stop experiment many times without pre-flattening the surface. When the rotation speed is larger than 30 rpm we have noticed the appearence of stratigraphy patterns [see Figure \ref{strat}]. For low rotation speeds, i.e. when the pattern is a close spiral, the ripple amplitude is so small that the ripples are destroyed by the rotation of the container. Therefore, no sedimentary structure can be observed.

In our experiment, it is possible to tune the ripple rate of migration and the rate of sand deposition by changing the rotation speed of the container. For very small values of the rotation speed ($\omega<30$ rpm) we are not able to observe sedimentary structure, as mentioned before. For medium rotation speed ($30<\omega<45$ rpm), the rate of migration is smaller than the rate of deposition. As a consequence, the ripples are observed as if they were deposed on each other (see left picture of Fig. \ref{strat}). This kind of structure is called a ``complete rippleform lamination'' \cite{hunter}. For large rotation speed ($\omega>45$ rpm), the rate of deposition is small while the rate of migration is large. The corresponding patterns are called truncated ripple-foreset crosslamination \cite{hunter} and are illustrated in the right part of Figure \ref{strat}. 

\begin{figure}[h]
\begin{center}
\includegraphics[width=8cm, angle=-90]{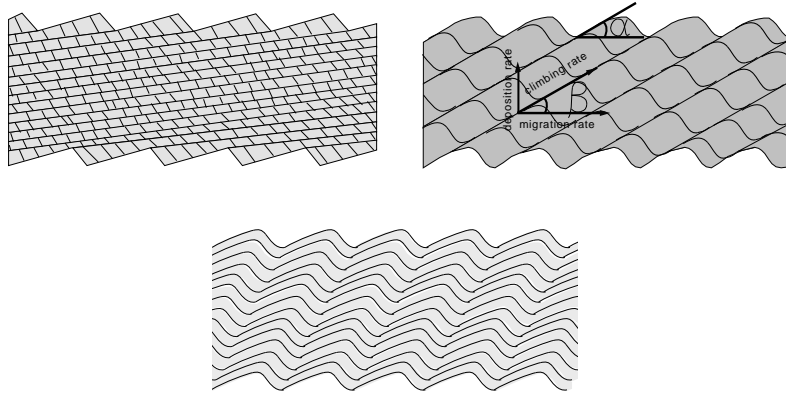}
\vskip -3cm
\caption{\small The angle of climbing $\alpha$ is the angle between the lamination direction and the horizontal one. The inclination angle of the upstream face of the ripples is called $\beta$. The climbing of the ripples creates a stratification pattern which depends on the difference between $\alpha$ and $\beta$: (top left) if $\alpha<\beta$, the sedimentary structure is said to be ``sub-critical'' and a truncated ripple-foreset crosslamination \cite{hunter} is observed; (top right) if $\alpha=\beta$, the sedimentary structure is ``critical'' and leads to a complete ripple-foreset crosslamination \cite{hunter}; (bottom) the case $\alpha>\beta$ corresponds to ``super-critical'' sedimentary structures and complete rippleform lamination \cite{hunter}.}\label{climb}
\end{center}
\end{figure}

In order to characterize the sedimentary structures, it is common \cite{hunter,makse} in geology to define a vector of translation [see Figure \ref{climb}]. The horizontal component of this vector corresponds to the rate of migration of the ripples, while the vertical component is the net rate of deposition. The {\it angle of climbing} $\alpha$ is defined as the angle between the climbing vector and the horizontal one. The inclination angle of the ripple upstream face is noted $\beta$ and has to be compared with $\alpha$. Depending on the values of $\alpha$ and $\beta$, different lamination patterns can be observed. If $\alpha<\beta$, the lamination structure is called ``sub-critically climbing'' \cite{hunter}, $\alpha=\beta$ causes a ``critical'' lamination \cite{hunter}, and for $\alpha>\beta$ the lamination is called ``super-critically climbing'' \cite{hunter}.

The left part of Figure \ref{strat} corresponds to a sedimentary structure created at medium ($30<\omega<45$ rpm) rotation speed. In this rotation speed range, the rate of migration is very small. Therefore the vector of translation is nearly vertical. The angle of climbing is thus $\alpha\approx 90^\circ$, and the sedimentary structure is ``super-critical''. For the right part of Figure \ref{strat}, the rotation speed $\omega\approx 55$ rpm. The angle values are $\alpha=19^\circ$ and $\beta=27^\circ$. The sedimentary structure is then ``sub-critical''. Such structures are observed for all values of $\omega>45$ rpm. 

We thus see that the rotation speed of the container changes the ``criticality'' of the sedimentary structures. One should finally note that the same kind of patterns has been observed in \cite{bagnold1,hunter} and modeled in \cite{makse} for aeolian sand ripples.  

\section{Summary}
We have presented an experimental study of hydraulic sand ripples formed with binary sand mixtures in a cylindrical container. The spiral shape of the ripple has been shown as the consequence of an instability in the fluid motion. We have shown the appearance of phase segregation and stratigraphy. The relative difference in size between both sand species has been highlighted as the physical parameter leading to the phase segregation. The rotation speed of the container has been shown to be a parameter leading to different lamination patterns.
\section*{Acknowledgements}
HC is financially supported by the FRIA, Brussels, Belgium. We thank M. Ausloos for fruitful discussions, and moreover, for commenting on the manuscript during its preparation.

\end{document}